\documentclass[preprint,pra,onecolumn]{revtex4}%
\usepackage{amsfonts}
\usepackage{amsmath}
\usepackage{amssymb}
\usepackage{graphicx}%
\begin{document}
\title{Manipulating quantum states with aspheric lenses }
\author{Zhi-Wei Wang }
\email{sdzzwzw@mail.ustc.edu.cn}
\author{Xi-Feng Ren }
\email{mldsb@mail.ustc.edu.cn}
\author{Yun-Feng Huang }
\author{Yong-Sheng Zhang }
\author{Guang-Can Guo }
\affiliation{Key Laboratory of Quantum Information, Department of Physics, University of
Science and Technology of China, Hefei 230026, People's Republic of China}

\begin{abstract}
We present an experimental demonstration to manipulate the width and position
of the down-converted beam waist. Our results can be used to engineer the
two-photon orbital angular momentum (OAM) entangled states (such as
concentrating OAM entangled states) and generate Hermite-Gaussian (HG) modes
entangled states.

PACS number(s): 03.67.Mn, 03.65.Ud, 42.50.Dv

\end{abstract}
\maketitle

\bigskip
\textbf{\ \ \ \ \ \ \ \ \ \ \ \ \ \ \ \ \ \ \ \ \ \ \ \ \ \ \ \ \ \ \ \ \ I.
INTRODUCTION}

Spontaneous parametric down-conversion (SPDC) generated in nonlinear crystals
is one of the most effective methods to obtain entangled photon
pairs\cite{Kwiat}, which forms the foundation of many applications in quantum
communication, quantum cryptography and quantum computation etc. Up to now the
theoretical discussions and experiments have mostly focused
on\ two-dimensional spaces, namely qubits. However, entanglement is also
embedded into the spatial modes of the two-photon states generated in SPDC.
Such entanglement can exist in infinite-dimensional Hilbert space so that it
has a promise to realize new-type protocols in quantum communication and
quantum cryptography\cite{2,3,4,5} compared with two-dimensional spaces
(qubits). It has been proven theoretically and\ experimentally that the
generated two-photon state\ from SPDC is entangled in orbital angular momentum
(OAM) \cite{Arnaut,Franke,Mair,Vaziri,Langford} which is related to spatial
modes of the two-photon states.

Normally the two-photon state at the output of the nonlinear crystal can be
written as a coherent superposition of eigenstates of the OAM
operator\cite{Franke}, while the amplitudes of each eigenstates are related to
the beam width of pump beam and signal (idler) beam\cite{11}. So we can
manipulate the final state by a proper selection of the beam width. Recently
Hermite-Gaussian (HG) modes entangled states have also been discussed\cite{HG}%
. It is underlined that if the condition $\omega_{0}/\omega_{p}\ll1$
($\omega_{0}$ is the width of the down-converted beam\ and $\omega_{p}$ is
that of the pump beam) is satisfied the HG modes of the signal and idler beam
can be considered as quasi-conserved. In this paper, we find that we can
manipulate the output state by properly choosing aspheric lenses for proper
parameters (the width and position of the down-converted beam waist) of the
down-converted photons can be selected with the help of lenses. So we can
realize the concentration of OAM entangled states. Similarly the new-type HG
modes entangled states can also be generated.

\textbf{\ \ \ \ \ \ \ \ \ \ \ \ \ \ \ \ \ \ \ \ \ \ \ \ \ \ \ \ \ \ \ \ \ \ \ II.
THE THEORY ANALYSIS}

In the process of SPDC, the pump light is incident on the nonlinear crystal
then the signal and idler photons can be generated with a small probability.
As mentioned above the spontaneous parametric down-conversion can generate
photon pairs entangled in OAM. It has also been shown that paraxial
Laguerre-Gaussian (LG) laser beams carry a well-defined OAM\cite{Allen}. The
beam $LG_{p}^{l}$ carries an OAM of $l\hbar$, where $l$ is referred to as the
winding number and $\left(  p+1\right)  $ is the number of radial nodes. So
the generated OAM entangled state can be expanded with LG modes\cite{11,10}%

\begin{equation}
|\psi\rangle=\sum_{l_{1},p_{1}}\sum_{l_{2},p_{2}}C_{p_{1},p_{2}}^{l_{1},l_{2}%
}|LG_{p_{1}}^{l_{1}};LG_{p_{2}}^{l_{2}}\rangle\text{,}%
\end{equation}
where $LG_{p_{1}}^{l_{1}}$\ corresponds to the mode of the signal beam and
$LG_{p_{2}}^{l_{2}}$ the mode of the idler beam. The weights of the quantum
superposition are given by $P_{p_{1},p_{2}}^{l_{1},l_{2}}$=$|C_{p_{1},p_{2}%
}^{l_{1},l_{2}}|^{2}$, which denoting the joint detection probability for
finding one photon in the signal mode $LG_{p_{1}}^{l_{1}}$ and the other in
the idler mode $LG_{p_{2}}^{l_{2}}$.

In most applications, we only consider $p_{1}=p_{2}=0$ for the sake of
simplicity. In this way, the two-photon state can be simplified as%
\begin{equation}
|\psi\rangle=\sum_{l_{1}=-\infty}^{\infty}\sum_{l_{2}=-\infty}^{\infty}%
C_{0,0}^{l_{1},l_{2}}|l_{1},l_{2}\rangle\text{.}%
\end{equation}

J. P. Torres has shown that $C_{0,0}^{l_{1},l_{2}}$ is related to the pump
beam width $\omega_{p}$ and the chosen beam width of the LG base $\omega_{0}$.
Generally speaking, $C_{0,0}^{l_{1},l_{2}}$ increases with $\omega_{p}$, but
there is an optimal value of $\omega_{0}$ for which the contribution of
$C_{0,0}^{0,0}$ is maximum\cite{11}.

To manipulate or utilize this OAM entangled state, we need to detect different
LG modes. The single-mode optical fibers and computer-generated holograms are
frequently used to solve this problem\cite{Mair,Vaziri,Vaziri and
Pan,Arlt,Leach}. In this paper, we just consider the detecting of $LG_{0}^{0}$
mode (the Gaussian mode), while the other LG modes can be done similarly with
the help of computer-generated holograms.

The detection efficiency for the single-mode fiber detecting photons in
$LG_{0}^{0}$ mode at the beam waist $\omega_{0}$ is given as\cite{Ren}
\begin{equation}
Q_{0}=\frac{\left(  \int\int\left(  LG_{0}^{0}\right)  ^{\ast}E\left(
\rho\right)  \rho d\rho d\varphi\right)  ^{2}}{\int\int\left(  LG_{0}%
^{0}\right)  ^{\ast}LG_{0}^{0}\rho d\rho d\varphi\int\int E\left(
\rho\right)  ^{\ast}E\left(  \rho\right)  \rho d\rho d\varphi}\text{,}%
\end{equation}
where
\begin{equation}
E\left(  \rho\right)  =E\left(  0\right)  \exp\left(  -\frac{\rho^{2}}%
{\omega_{f}^{2}}\right)  \text{,}%
\end{equation}
$E\left(  0\right)  $ denoting the amplitude of the field at the fiber centre
and $d=2\omega_{f}$ is the mode field diameter of the fiber.

For the sake of simplicity, we assume that the maximal detection efficiency
corresponds to the case that the beam waist of the down-converted light is
incident on the input surface of the coupling lens of single-mode fiber. At
this time, $\omega_{0}^{\prime}$\ is the beam width and $z^{\prime}$ is the
distance from the lens to the beam waist after the lens. By virtue of
transformation of the lens as shown in Figure 1:
\begin{equation}
\omega_{0}^{\prime2}=\frac{\omega_{0}^{2}}{\left(  1-\frac{z}{f}\right)
^{2}+\frac{\pi^{2}\omega_{0}^{4}}{\lambda^{2}f^{2}}}%
\end{equation}

\begin{equation}
z^{\prime}=\left[  1-\frac{\left(  1-\frac{z}{f}\right)  }{\left(  1-\frac
{z}{f}\right)  ^{2}+\frac{\pi^{2}\omega_{0}^{4}}{\lambda^{2}f^{2}}}\right]
f\text{.}%
\end{equation}
We can calculate the parameters $\left(  \omega_{0},z\right)  $ of the
down-converted beam before the lens if we know the values of $\omega
_{0}^{\prime}$ and $z^{\prime}$.

In the experiment, we change the position of the coupling lens, while the
positions of the BBO crystal and the single-mode fiber are stable. The
detection efficiency $Q_{0}$, substituted by the count rate $R$, is changing
when we change the position of the lens. From the experiment data, the optimal
values of $\omega_{0}^{\prime}$ and $z^{\prime}$ can be obtained by means of
\textquotedblleft Curve Fitting\textquotedblright, which is the least squares
fit program in OriginPro 7.0. Then from Eqs. (5) and (6), we can obtain the
values of $\omega_{0}$ and $z$. While the value of the width of the pump beam
$\omega_{p}$ can be easily manipulated by adding another lens in the pump
beam. Thus we realize the manipulation of the values of $\omega_{p}$ and
$\omega_{0}$.

\ \ \ \ \ \ \ \ \ \ \ \ \ \ \ \ \textbf{III. THE EXPERIMENT CONFIGURATION AND
ANALYSIS}

The experiment configuration is shown in Figure 2. A BBO\ (beta barium borate)
crystal (thickness $1$ $mm$) is illuminated by a quasimonochromatic
argon-ion\ laser pump beam propagating in the $z$ direction at the wavelength
$\lambda_{p}=351.1$ $nm$. In the pump light path, an aspheric lens A
$(f_{A}=500$ $mm)$ is placed to change the pump beam width $\omega_{p}$ on the
input surface of the crystal. After an appropriate choice of the phase
matching angle, the down-converted photons are produced in type I SPDC at
degenerated wavelength of $702.2$ $nm$ at an angle of $6^{\circ}$\ off the
pump light direction. In the path of the signal photons, another aspheric lens
B is placed to focus the photons into the detector. Before the detector are an
interference filter (bandwidth $4$ $nm$) and a fixed coupling lens (it is not
shown in Figure 2) used to couple the down-converted photons into the
single-mode fiber.

In Figure 3 the horizontal axis is the distance from the detector to lens B
and the vertical axis is the count rate for the signal path. In the
experiment, we scan the position of lens B and write down the corresponding
count rate with the position of the crystal and the detector fixed (the
distance from the detector to the BBO crystal is $852$ $mm$). The dots are the
experiment data of single counts in the signal path, while the curves are
based on the \textquotedblleft Curve Fitting\textquotedblright\ with the dots.
It can be seen that the experiment data fit the simulated curves very well
near the peak. But far away from the peak, there is some deviation. Such
deviation is probably due to the fact that when lens B is far away from the
position of the peak, the light incident on the input surface of the lens
group has a larger scattering angular. At this time, the Eq. (3) may be not
appropriate. The values of $\left(  \omega_{0}^{\prime},z^{\prime}\right)  $
come from the \textquotedblleft Curve Fitting\textquotedblright\ and the
values of $\left(  \omega_{0},z\right)  $ are calculated from the ones of
$\left(  \omega_{0}^{\prime},z^{\prime}\right)  $\ according to Eqs. (5) and (6).

In the experiment, first the role of $\omega_{p}$ played in the detection of
the spatial mode\ is investigated. We change the value of $\omega_{p}$ by
varying the position of lens A. In Figure 3(a), (b), (c), $\left(
f_{B}=100\text{ }mm\right)  $, the relations $\omega_{p,a}<\omega_{p,b}%
<\omega_{p,c}$ can be derived by proper position of lens A. From the three
charts, we find the count rate gets larger as $\omega_{p}$ becomes smaller.
For further justification of this result, in Figure 3(d),(e), lens B with the
focus of $200$ $mm$ is used. The same result is derived as above. The result
is in good accords with the theory put forward by Torres et al. that the count
rate of $LG_{0}^{0}$ mode can be made larger by decreasing the pump beam
width\cite{11}.

Comparing the case of $f_{B}=200$ $mm$ with that of $f_{B}=100$ $mm$, one can
find that the similar values of $\left(  \omega_{0},z\right)  $ are got for
the same lens B. For example, in the case of $f_{B}=100$ $mm$, we get the
values of $\left(  \omega_{0},z\right)  $ $(0.024,115.1)$, $(0.026,114.7)$ and
$(0.029,114.5)$ respectively, and when $f_{B}=200$ $mm$, the values of
$\left(  \omega_{0},z\right)  $ are $\left(  0.083,281.6\right)  $, $\left(
0.078,282.6\right)  $ respectively. We can also find that the parameters with
regard to different focus of lens B are obviously distinct from each other,
which means that lens B is able to \textit{select} the values of $\left(
\omega_{0},z\right)  $. Since different values of $\left(  \omega
_{0},z\right)  $ represent different spatial distribution of down-converted
light field, we can draw the conclusion that different lens B can
\textit{select} different spatial distribution of down-converted light field.

From the analysis above, we can vary the values of $\omega_{p}$ and
$\omega_{0}$ by means of proper operation on lenses. As the amplitude
$C_{p_{1},p_{2}}^{l_{1},l_{2}}$ in Eq. (1) is related to $\omega_{p}$ and
$\omega_{0}$, we can engineer the state of Eq. (1) by selecting the focus and
position of lens. So we can obtain the maximal OAM entangled states using this
method with the technology available. In the same way, partial entangled
states can also be derived for they both have many important
applications\cite{Horodechi,Mozes}. Similarly, the condition $\omega
_{0}/\omega_{p}\ll1$ can be satisfied easily, allowing to produce the HG modes
entangled states\cite{HG}.

\ \ \ \ \ \ \ \ \ \ \ \ \ \ \ \ \ \ \ \ \ \ \ \ \ \ \ \ \ \ \ \ \ \ \ \ \ \ \ \textbf{IV.
CONCLUSIONS}\ \ \ \ \ \ \ \ \ \ \ \ \ \ \ \ \ \ \ \ \ \ \ \ \ \ \ \ \ \ \ \ \ \ \ \ \ 

In conclusion, we find that using detection efficiency of the down-converted
photons and \textquotedblleft Curve Fitting\textquotedblright\ function, we
can acquire the parameters of the down-converted light field of SPDC. We can
manipulate the state in Eq. (1) by selecting the focus and position of lenses.
The condition to generate HG modes entangled states can also be satisfied
using this method. Additionally our conclusion is useful to increase detection
efficiency of the down-converted photons.

\begin{center}
\textbf{ACKNOWLEDGMENTS}
\end{center}

This work was funded by the National Fundamental Research Program
(2001CB309300), National Natural Science Foundation of China(10304017,
10404027), the Innovation Funds from Chinese Academy of Sciences.

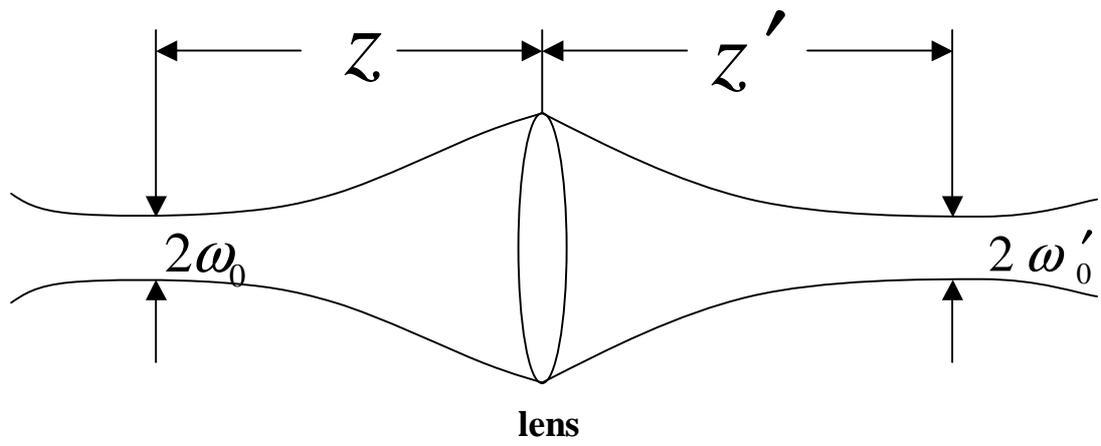
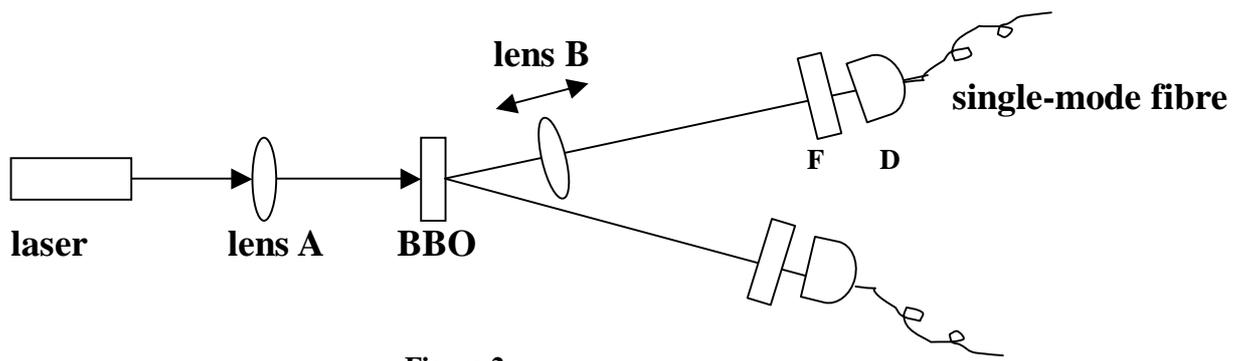
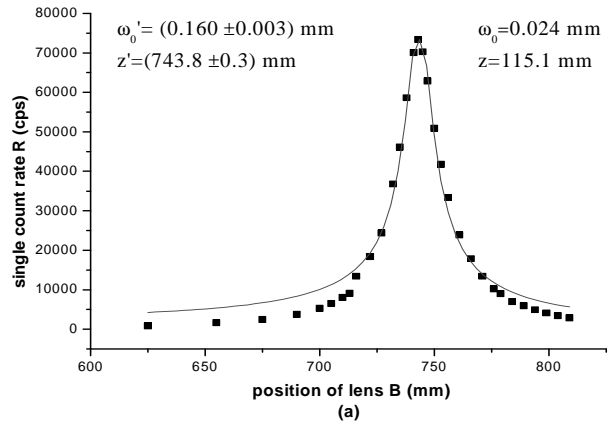
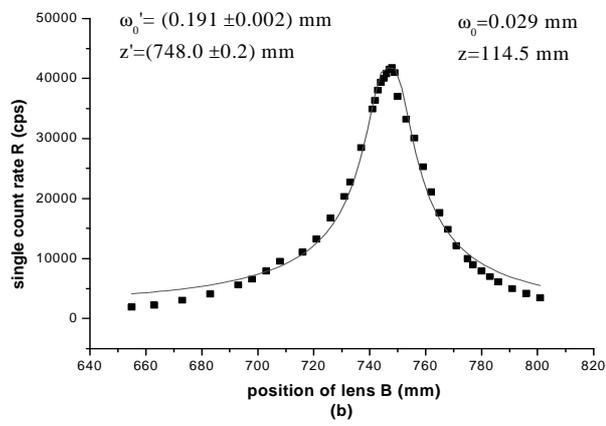
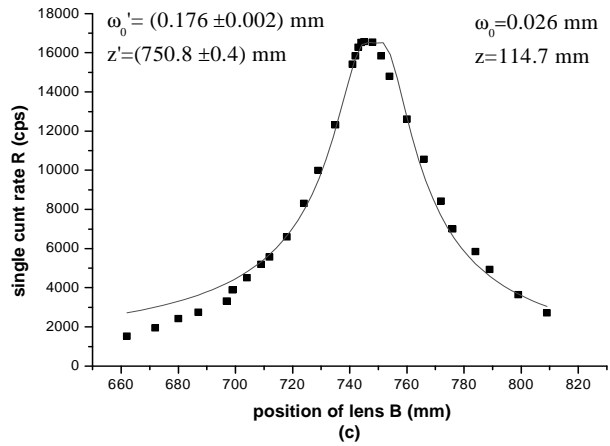
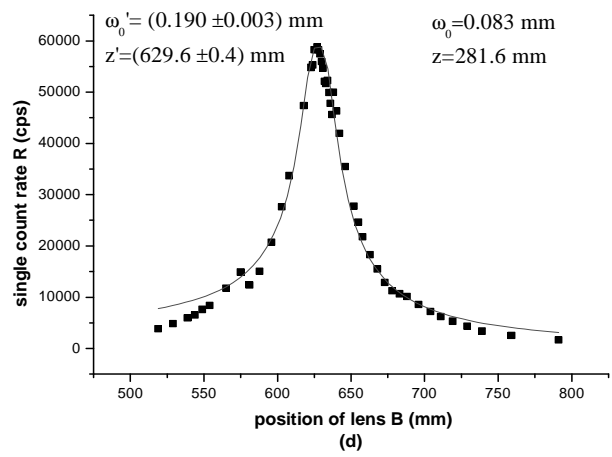
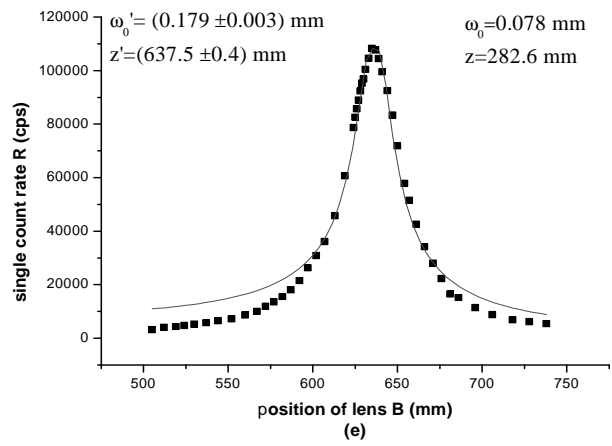
\end{document}